\documentclass[%
 reprint,
 amsmath,amssymb,
 aps,
]{revtex4-1}

\usepackage{graphicx}
\usepackage{dcolumn}
\usepackage{bm}

\begin{document}

\preprint{APS/123-QED}

\title{Temperature and lifetime measurements in the SSX wind tunnel}

\author{M. Kaur}
\author{K. D. Gelber}%
\author{A. D. Light}%
\author{M. R. Brown}%
\affiliation{%
 Swarthmore College, Department of Physics and Astronomy, Swarthmore, PA 19081\\
}%

\date{\today}%

\begin{abstract}
We describe electron temperature measurements in the SSX MHD wind tunnel using two different methods. First, we estimate $T_e$ along a chord by measuring the ratio of the $C_{III}~97.7~nm$ to $C_{IV}~155~nm$ line intensities using a vacuum ultraviolet monochrometer. Second, we record a biasing scan to a double Langmuir probe to obtain a local measurement of $T_e$. The aim of these studies is to increase the Taylor state lifetime, primarily by increasing the electron temperature. Also, a model is proposed to predict magnetic lifetime of relaxed states and is found of predict the lifetime satisfactorily. Furthermore, we find that proton cooling can be explained by equilibration with the electrons.
\end{abstract}

\maketitle


\section{\label{sec:Intro}Introduction}

Long magnetic lifetimes are essential for magnetic fusion energy schemes, particularly magneto-inertial fusion (MIF) \cite{Lindemuth95,Binderbauer10,Degnan13,Wurden16}.  Since magnetic lifetimes scale as $L/R$ (inductance by resistance), and since plasma resistivity scales as $T_e^{-3/2}$, high electron temperatures are essential for these schemes \cite{Peacock69}.  Often, MIF targets are imploded on mechanical, i.e., slow, time scales \cite{Laberge09} so they need to remain stable with sustained magnetic flux for at least $100~\mu s$.

We have performed experiments on the SSX plasma wind tunnel \cite{BrownJPP,BrownPSST} with an eye towards increasing the magnetic lifetime of our Taylor state plasmas \cite{Taylor74,Taylor86,Gray13,Cothran09}.  We propose a simple model for the magnetic lifetime based on Spitzer resistivity and the force-free eigenvalue.  The model satisfactorily predicts the magnetic e-folding time using the measured value for $T_e$.  In addition, we measure $T_i$ with ion Doppler spectroscopy and find the warm protons cool on the electrons.

In section \ref{sec:SSX}, we discuss the SSX plasma wind tunnel and the generation of relaxed Taylor state plasmas. In sections \ref{sec:VUV} and \ref{sec:DLP}, we review the operation of our key diagnostics: the vacuum ultraviolet (VUV) monochrometer and double Langmuir probe.  In section \ref{sec:Results} we present our results.

\section{\label{sec:SSX}SSX plasma wind tunnel}

\subsection{\label{sec:SSX1}SSX device}

\begin{figure*}[!h]
\begin{center}
\includegraphics[width=0.95\textwidth]{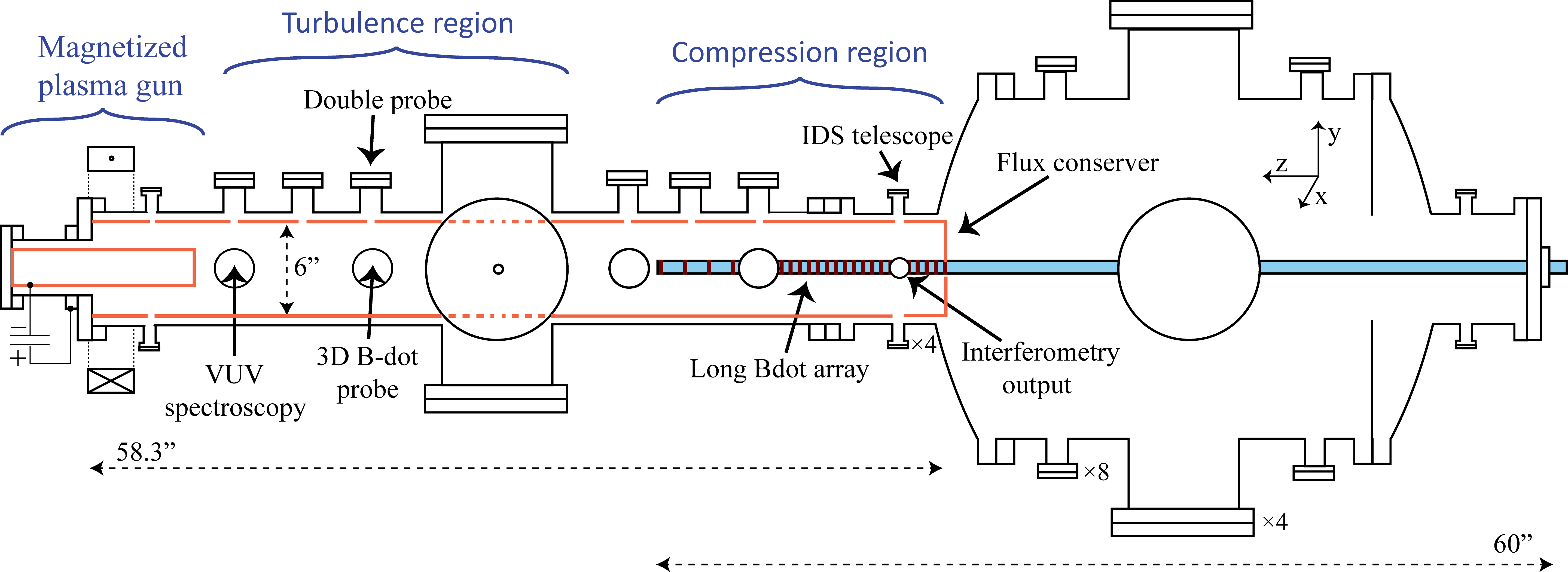}
\end{center}
\caption{Schematic of the SSX device in the wind tunnel configuration. The entire device is divided into three main sections: $(i)$ plasma source region, $(ii)$ turbulence studies region and $(iii)$ the compression region. In the first region, a magnetized coaxial plasma gun is installed which generates fully-ionized, magnetized plasma. In the turbulence region, VUV spectroscopy is installed $5~cm$ away from the source for $T_e$ measurements. In the same region at $25~cm$ away from the gun, a double Langmuir probe and a co-located $\dot{B}$ probe are installed to measure $T_e$, $n$ and three orthogonal components of the magnetic field. In the compression region, a long $\dot{B}$ probe array is aligned axially to measure magnetic field structure and time of flight velocity. In addition, an ion Doppler spectroscopy and HeNe laser interferometry are used for measuring the ion temperature and plasma density, respectively at a distance of $124~cm$ away from the gun.}
\label{schematic} 
\end{figure*}

The Swarthmore Spheromak Experiment (SSX) plasma wind tunnel configuration features a $L \cong 1.5~m$ long, high vacuum chamber in which we generate $n \ge 10^{15}~cm^{-3}, T \ge 20~eV, B \le 0.5~T$ hydrogen plasmas \cite{BrownJPP,BrownPSST} (See Figure \ref{schematic}). The entire set-up is divided into three main sections: $(i)$ plasma source region, $(ii)$ turbulence region and $(iii)$ the compression region. In the first region, a magnetized coaxial plasma gun is installed which generates fully-ionized, magnetized plasma. 

Plasmas are accelerated to high velocity ($\cong 50~km/s$) by ${\bf J \times B}$ forces in the gun with discharge currents up to $100~kA$ and injected into a highly evacuated, field-free target volume called a flux conserver. The flux conserver is cylindrical in shape and bounded by a thick, highly conducting copper shell ($r = 0.08~m$). The inner, plasma-facing surface of the flux conserver is coated with tungsten.

The plasma ejected out of the gun relaxes to a twisted Taylor state through turbulence. During this relaxation phase, the turbulence studies can be carried out.  
The VUV spectroscopy is installed $5~cm$ away from the gun (in the turbulence region) for electron temperature ($T_e$) measurements. $T_e$ measurements are utilized for determining the confinement time and lifetime studies. 

\begin{figure}[!h]
\begin{center}
\includegraphics[height=3.5in]{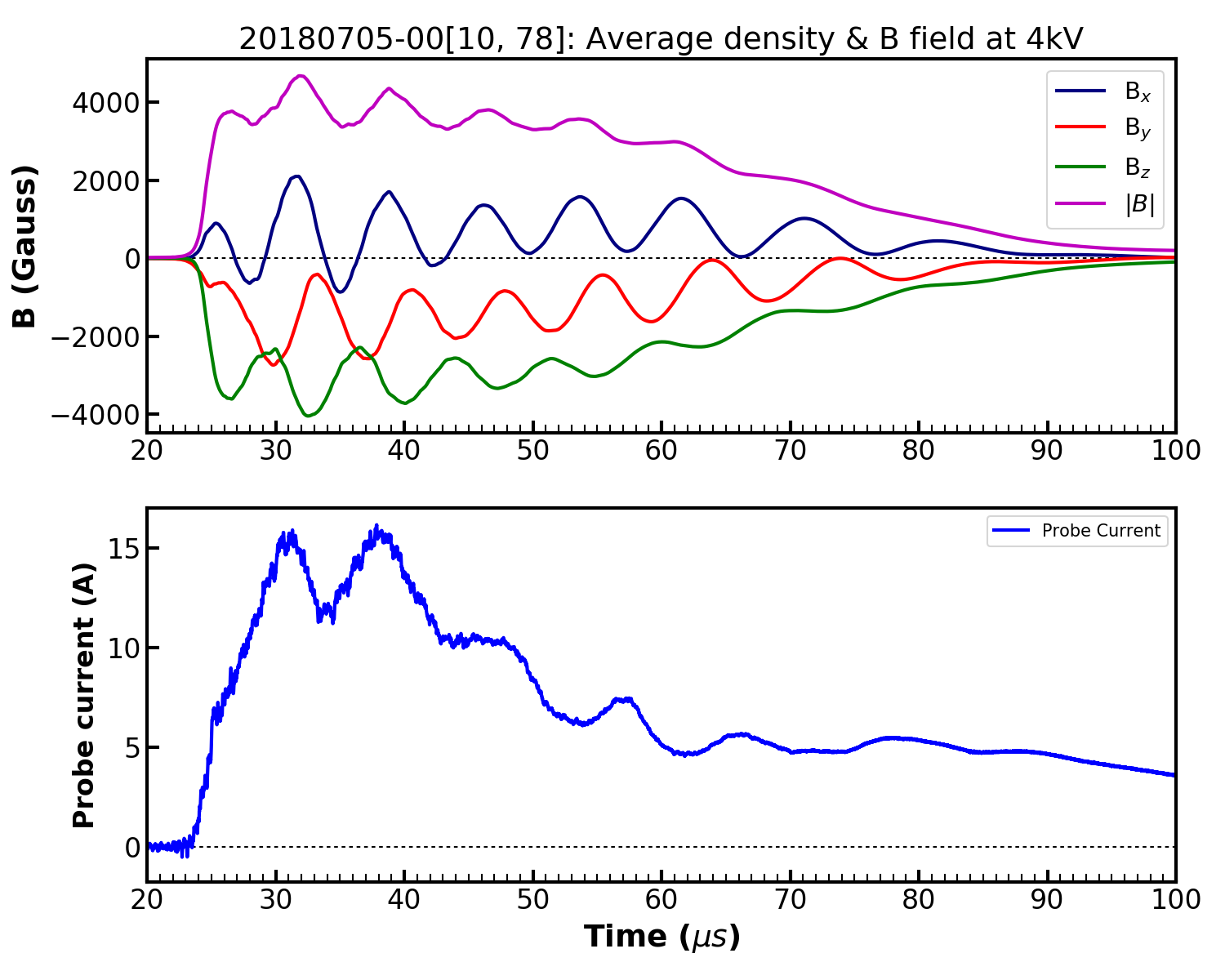}
\end{center}
\caption{Typical SSX wind tunnel data. Magnetic field (above, three components and $|B|$), double Langmuir probe ion saturation current (below).  Magnetic field data is a 69 shot ensemble average.  Measurement is made at the location shown in Figure \ref{schematic}.}
\label{BNdata} 
\end{figure}

In the same region at $25~cm$ away from the gun, a double Langmuir probe and a co-located $\dot{B}$ probe are installed to measure $T_e$, plasma density ($n$) and three orthogonal components of the magnetic field (Figure \ref{BNdata}). Data from these probes is used for carrying out correlation studies on the fluctuations present in density and magnetic field (outside the scope of the present work).

In the compression region, a long $\dot{B}$ probe array is aligned axially to measure magnetic field structure and time of flight velocity. $124~cm$ away from the gun, an ion Doppler spectrometer, HeNe laser interferometer, and another double probe are used for measuring (respectively) the ion temperature, a line-averaged plasma density, local electron temperature, and local plasma density (Figure \ref{density}).
\bigskip

\subsection{\label{sec:SSX2}Taylor states}

Relaxed Taylor states are the minimum energy states of magnetohydrodynamics.  Taylor state fields obey the force-free eigenvalue condition: $\nabla \times B = \lambda B$ \cite{Taylor74,Taylor86}.  Where the eigenvalue $\lambda$ can be written: $$\lambda = \frac{\nabla \times B}{B} = \frac{\mu_0 J}{B} = \frac{\mu_0 I}{\Phi}$$ where we have used Ampere's law, and integrated over an area.  The definition of inductance is: $\Phi = L I$ where $\Phi$ is the flux.  So we have: $$\lambda = \frac{\mu_0 I}{L I} = \frac{\mu_0}{L}$$ and a simple formula for inductance: $L = \mu_0/\lambda$.  This is satisfying since we know that Taylor states seek the minimum $\lambda$, so that means they seek to maximize their inductance.  The infinite Taylor ground state has $\lambda r = 3.11$ (or $\lambda = 40~m^{-1}$, for $r = 0.08~m$) \cite{Gray13,Cothran09}.
%
%
%
%
%
%
%
%

Roughly, the resistance is the resistivity $\eta$ times some length, divided by some area.  So $R$ has the units of $\eta/\ell$ or $\lambda \eta$, since the characteristic length scale of the Taylor state is defined by its eigenvalue.  This suggests a compact formula for estimating the lifetime: 
\begin{equation}
\tau = \frac{L}{R} = \frac{\mu_0}{\lambda^2 \eta}.  
\end{equation}
For $T_e = 10~eV$, we predict $\tau = 50~\mu s$ using this expression. 

\bigskip

\subsection{\label{sec:SSX3}Equilibration of proton and electron temperatures}
Cooling of the proton component of the SSX plasma due to collisions with the electrons can be modeled with a simple Newton cooling-type equation:
\begin{equation}
\frac{dT_p}{dt} = \nu_{pe} (T_e - T_p)  
\end{equation}
where $T_p$ is the proton temperature, and $T_e$ is the electron temperature, The proton-electron equilibration rate \cite{NRL} can be written: $$\nu_{pe} = 3.24 \times 10^{-9} \frac{n_e~ln\Lambda}{(T_e + T_i/1836)^{3/2}}$$ where $ln \Lambda$ is the Coulomb logarithm, temperatures are measured in $eV$ and densities in $cm^{-3}$.  Assuming a nearly constant electron temperature and if $T_e$ and $T_p$ are not too different, equilibration times are several $\mu s$ for $T_e = 10~eV$, $n_e \ge 10^{15}~cm^{-3}$, and $ln \Lambda = 8.4$.  

\section{\label{sec:VUV}VUV spectroscopy}

\subsection{\label{sec:VUV1}Background}

The first Vacuum Ultraviolet (VUV) spectrograph was built by Schumann in 1893  \cite{Milazzo69}. In the decades since then, the accuracy and ability to measure radiation in the ultraviolet region, the range from $210~nm$, down to about $1~nm$, has increased dramatically \cite{Milazzo69}. Given that both glass and air are opaque in this region, the measurements  - including the transport, diffraction, and detection of UV photons - must be performed either under vacuum, or in a gas that is transparent in the ultraviolet region. Although spectroscopy has the disadvantage of making line averaged rather than localized measurements, it is noninvasive and does not perturb the plasma. 

In our experiment, the ultraviolet radiation emitted by the plasma is transmitted under vacuum to our VUV monochromator, where the photons strike a magnesium-fluoride coated diffraction grating, then pass into a a sodium salicylate scintillator which fluoresces at $420~nm$. The $420~nm$ photons pass through a glass vacuum port and be detected by a photomultiplier tube (PMT), which produces a trace of the diffracted line intensity versus time. \cite{McPherson}

\subsection{\label{sec:VUV2}SSX VUV system}

\begin{figure}[!h]
\begin{center}
\includegraphics[height=3.5in]{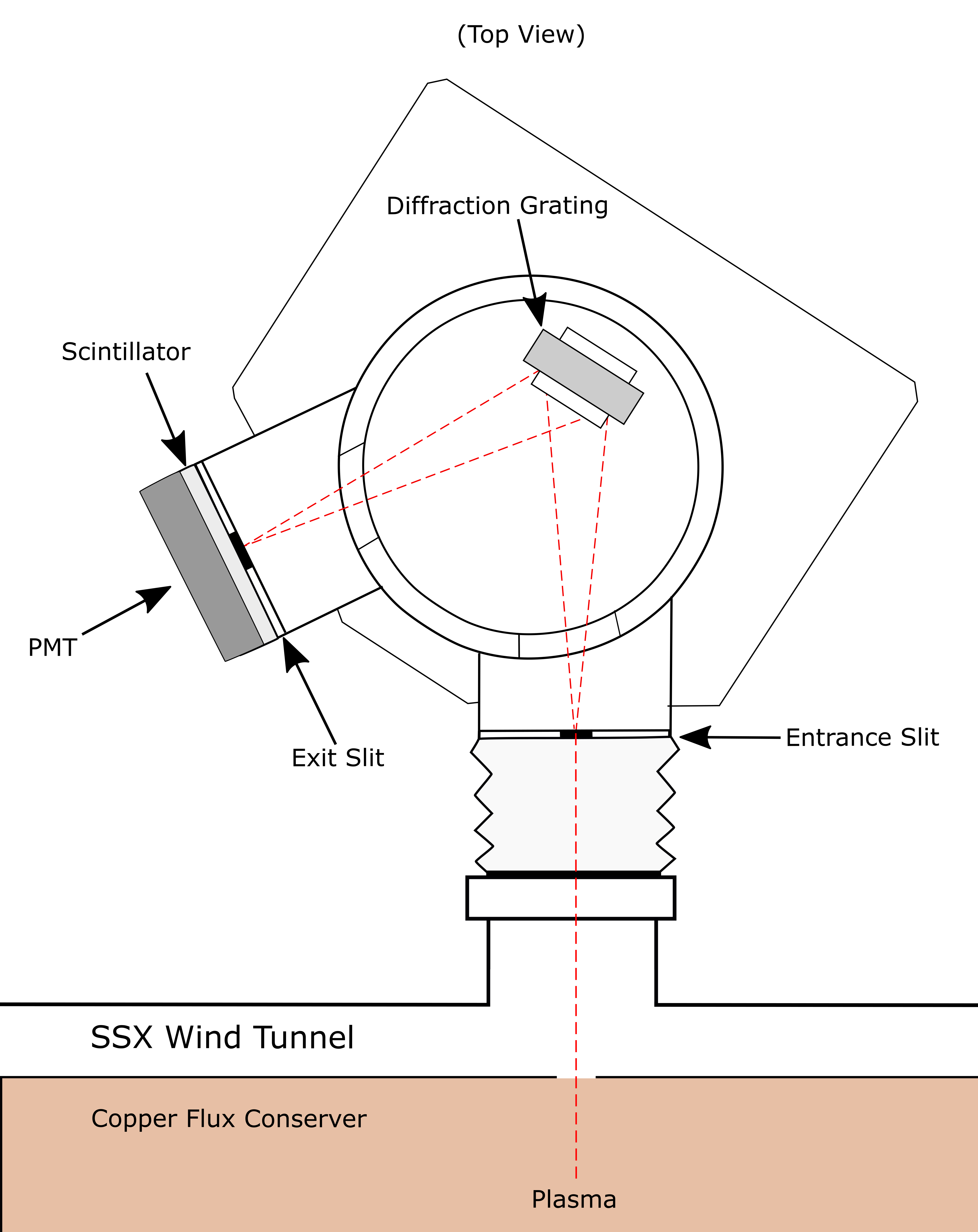}
\end{center}
\caption{Schematic of VUV monochromator. Radiation emitted by the plasma passes thought a hole in the copper flux conserver, then though a flange into the monochromator, all under vacuum. The diffraction grating selects and refocuses a narrow band of light centered at a selected wavelength. The photons strike a scintillator, which fluoresces and the signal is detected and amplified by the PMT.}
\label{VUV} 
\end{figure}

We measure the electron temperature along a $0.16~m$ chord by taking the ratio of $C_{III}~97.7~nm$ ($^1P_1 \rightarrow ^1S_0$) to $C_{IV}~155~nm$ ($^2P_{1/2.3/2} \rightarrow ^2S_{1/2}$) line intensities \cite{Chaplin09} using a McPherson 234 M8 $0.2~m$ vacuum ultraviolet monochrometer (see Figure \ref{VUV}). Both the entrance and exit slits of the monochrometer are adjustable. The entrance slit controls the number of photons into the monochromator, while the exit slit can be used to control the spectral resolution. 

We use a $1200/mm$ grating so the dispersion of our $0.2~m$ monochrometer is $$\frac{\Delta \lambda}{\lambda} = sin \theta \approx \frac{1/1200~mm}{200~mm} = 4~nm/mm.$$  Resolution is dispersion times detector size so $$\mathcal{R} = 4~nm/mm \times 0.125~mm = 0.5~nm$$ if we use $125~\mu m$ exit slits. This is the optimum resolution of our monochrometer. 

\begin{figure}[!h]
\begin{center}
\includegraphics[height=3.5in]{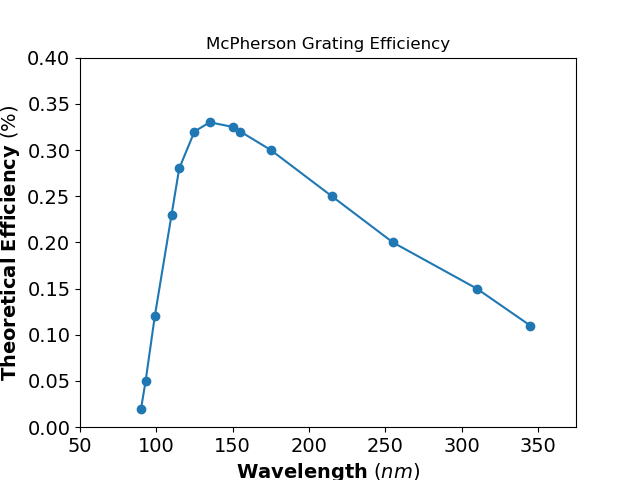}
\end{center}
\caption{Grating efficiency of McPherson 1200 grooves/mm $Al+MgF_2$ grating.  We operate at $97~nm$ and $155~nm$ where the grating efficiency differs by a factor of 3.}
\label{MgF2} 
\end{figure}

We had previously developed a calibration curve for the VUV monochrometer at visible wavelengths \citep{Chaplin09}.  During SSX experiments, we were able to scan through the $97.7~nm$ and $155~nm$ lines to ensure we were at the center wavelength, and confirm the original calibration.  The $MgF_2$ coated aluminum grating ($1200/mm$) does not have flat efficiency over our wavelength range (Figure \ref{MgF2}); the grating efficiency differs by a factor of 3 for the $97.7~nm$ and $155~nm$ lines.
%
%
%
%

\begin{figure}[!h]
\begin{center}
\includegraphics[height=3.5in]{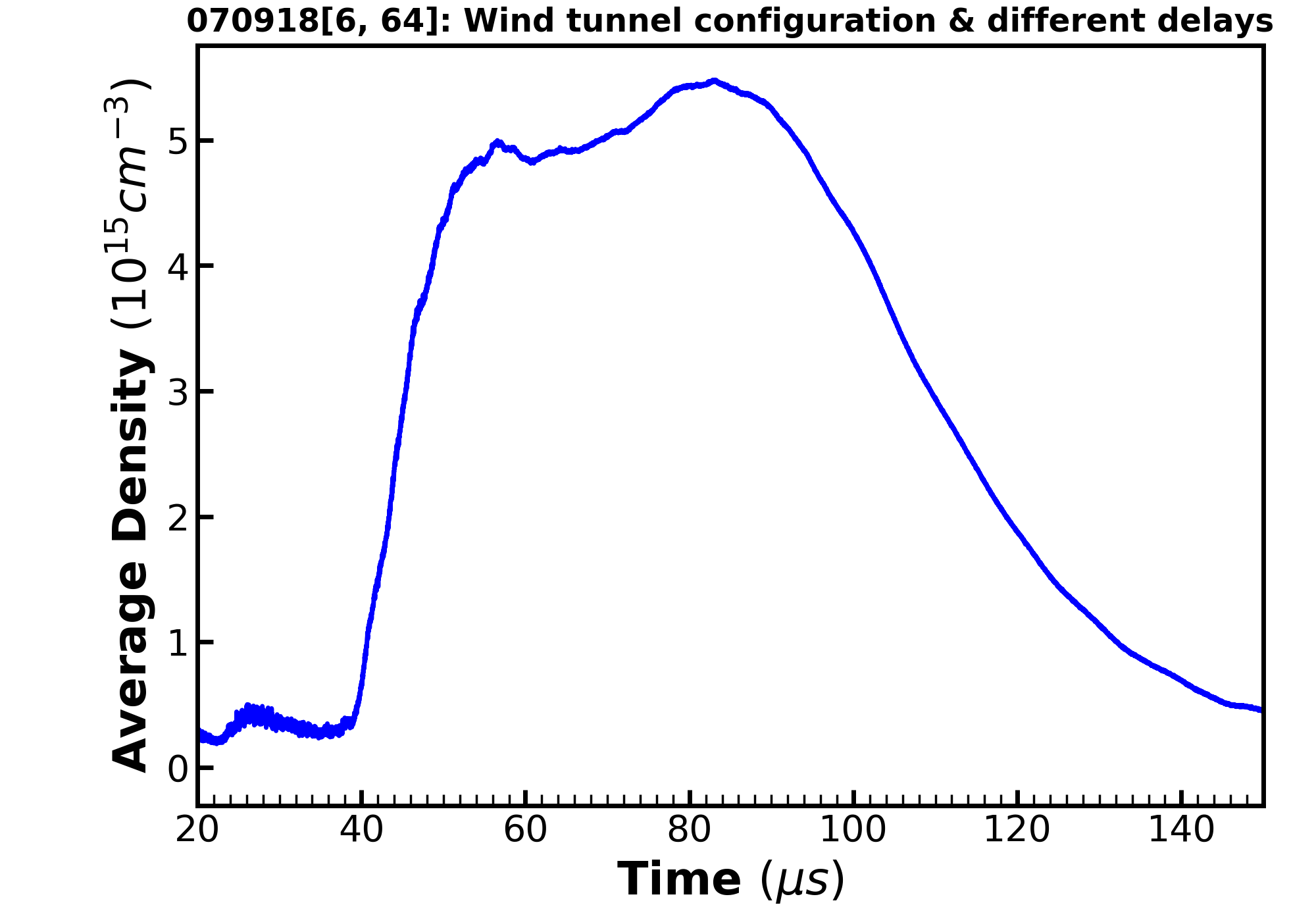}
\end{center}
\caption{Time trace of average electron density obtained from HeNe laser interferometry measured $124~cm$ away from the gun from $58$ plasma shots.}
\label{density} 
\end{figure}

\begin{figure}[!h]
\begin{center}
\includegraphics[height=3.5in]{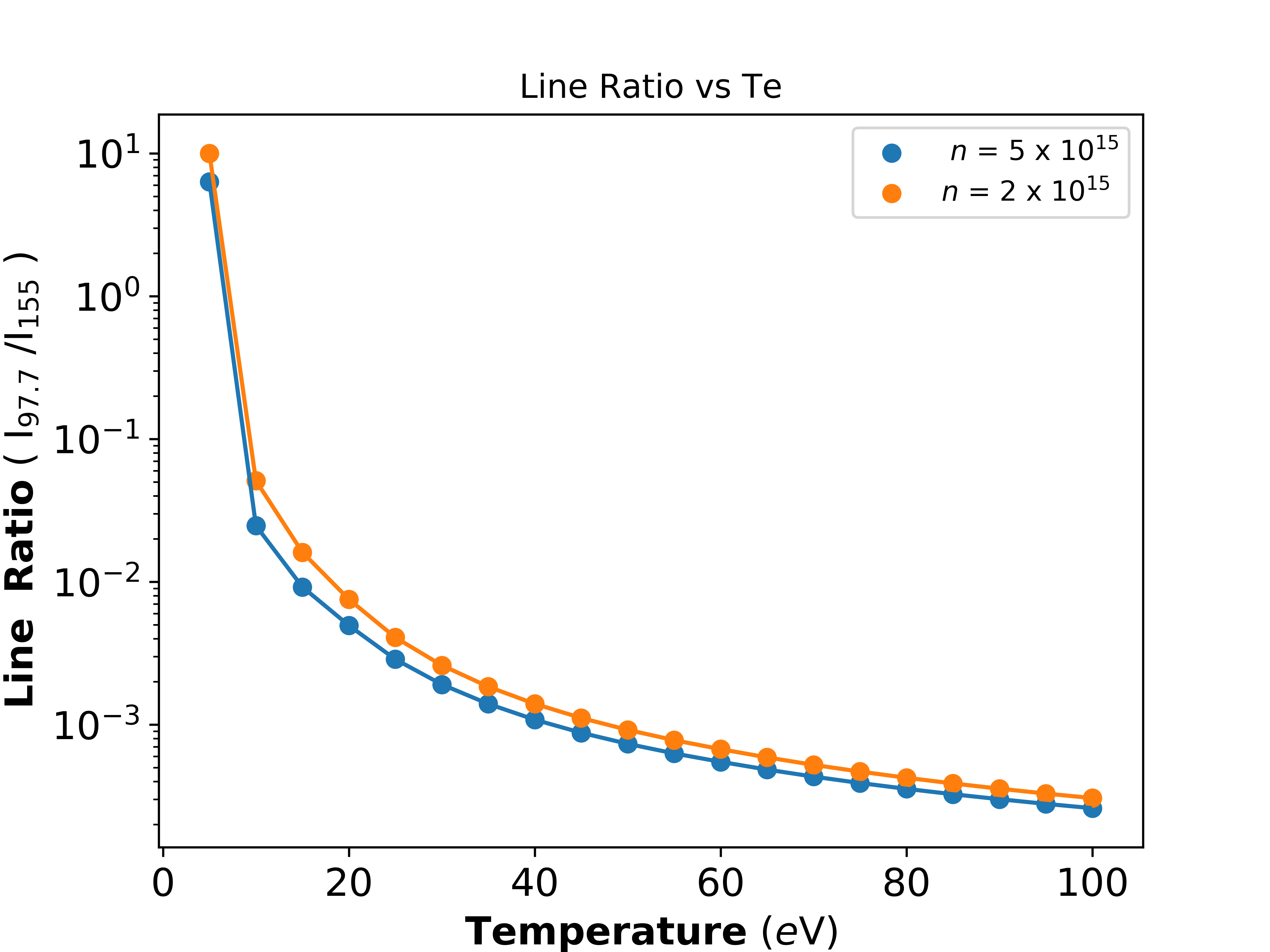}
\end{center}
\caption{Calibration Curve. Intensity ratio of the $97.7~nm~C_{III}$ to the $155~nm~C_{IV}$ as a function of $T_e$ in electron volts, at two densities. Ratios were calculated using a non-LTE excitation kinematics code (PrismSpec) at several densities and temperatures and interpolated between. \cite{Chaplin09}.}
\label{calib} 
\end{figure}

The $C_{III}$ and $C_{IV}$ line intensities have different temperature dependences.  Using the measured density of the plasma (Figure \ref{density}), we can relate the measured ratio of $C_{III}~97.7~nm$ and $C_{IV}~155~nm$ intensity to the electron temperature. This data was calculated using a non-LTE excitation kinematics code (PrismSPECT) at several densities and temperatures and interpolated smoothly between calculated points \cite{Chaplin09} (Figure \ref{calib}). The corrected photocurrents are smoothed over a .25 {\it $\mu$s} window, and standard error is calculated. Then the lines are divided and then the ratio at each point in time is converted into a temperature, using Figure \ref{calib}. 
%
%
%
%

\begin{figure}[!h]
\begin{center}
\includegraphics[height=3.5in]{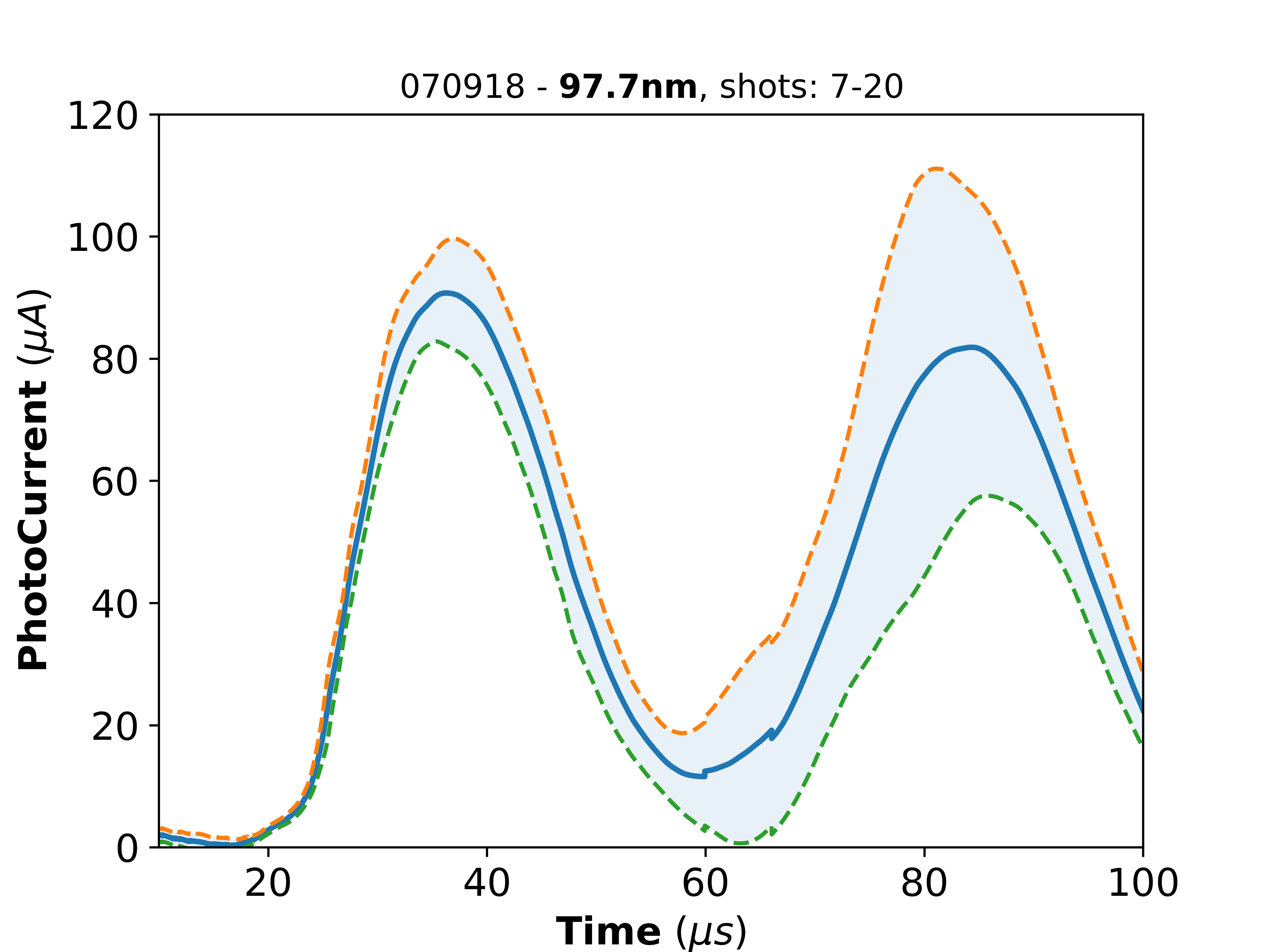}
\includegraphics[height=3.5in]{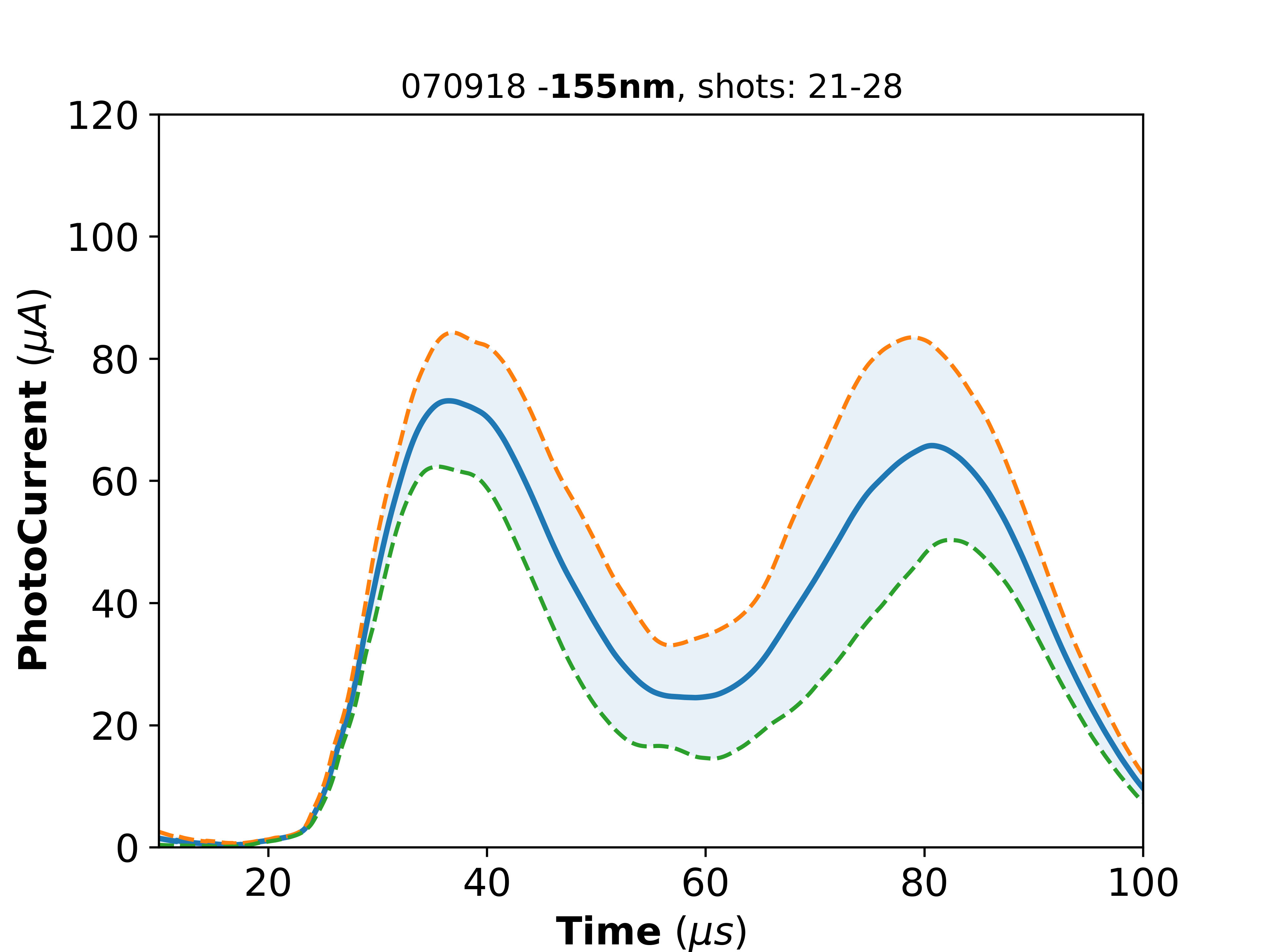}
\end{center}
\caption{Typical photocurrent measured at $97.7~nm$ (top) and $155~nm$ (bottom) using VUV spectroscopy. The $97.7~nm$ line corresponds $C_{III}$ transitions (14 shot average) and the $155~nm$ line corresponds to $C_{IV}$ transitions (8 shot average). Error bars reflect the standard error.  We divide the $97~nm$ signal by the $155~nm$ signal to measure $T_e$.}
\label{photocurrent} 
\end{figure}

\begin{figure}[!h]
\begin{center}
\includegraphics[height=3.5in]{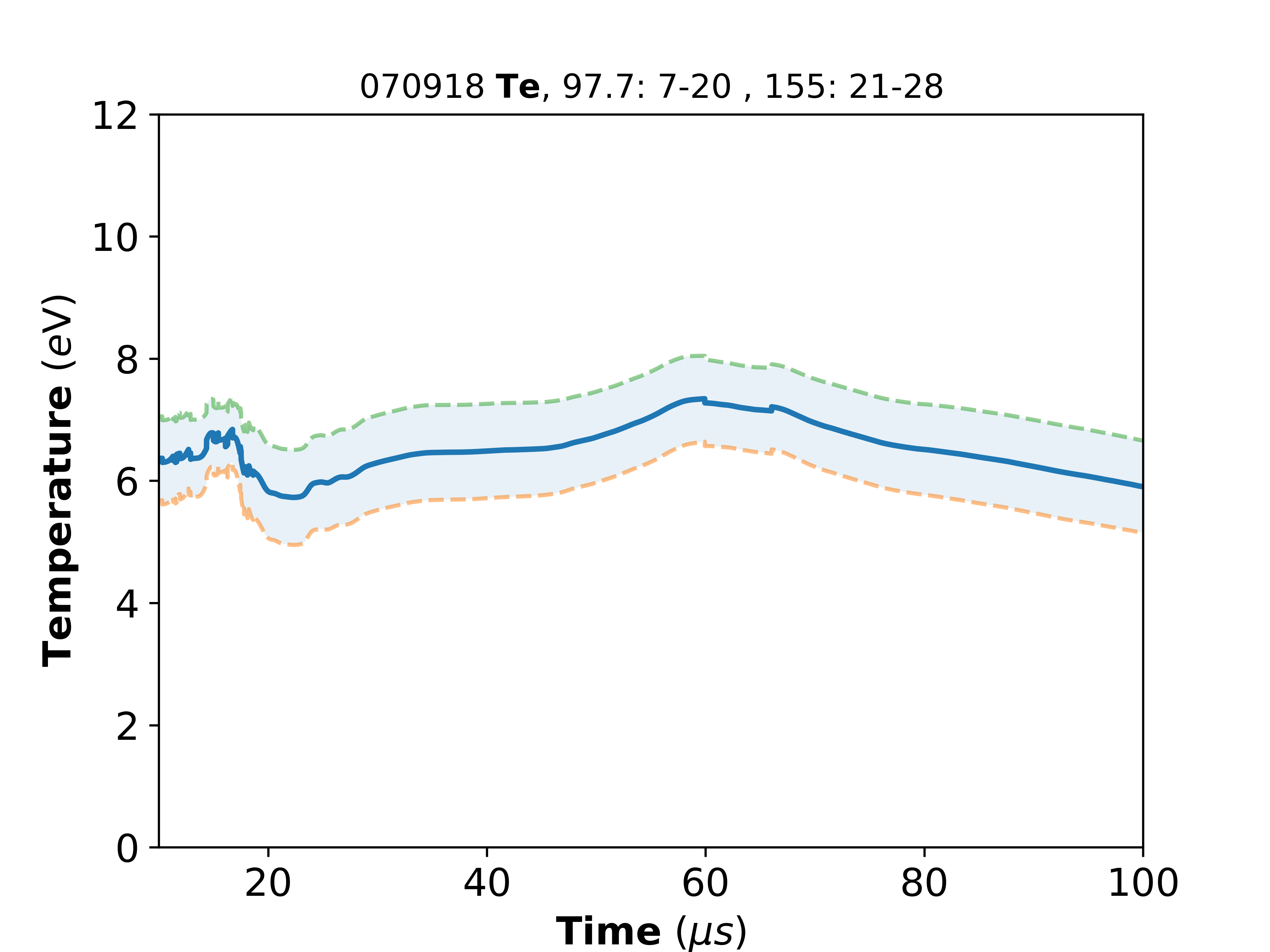}
\end{center}
\caption{Electron temperature. $T_e$ was measured by averaging the ratio between the intensities of CIII and CIV photons, and then converted to temperature using the calibration curve, assuming a density of $5 \times 10^{15}$ on Figure \ref{calib}.  Error bars are propagated from Figure \ref{photocurrent}.}
\label{TeVUV} 
\end{figure}

In Figure \ref{photocurrent} we present the corrected photocurrent for both $C_{III}~97.7~nm$ and $C_{IV}~155~nm$.  Approximately 10 shots are averaged to obtain a good average with acceptable errors.  In Figure \ref{TeVUV} we present $T_e(t)$ for the data in Figure \ref{photocurrent}, using Figure \ref{calib} and a fixed density of   $5 \times 10^{15}~cm^{-3}$ (see Figure \ref{density}) to project the line ratio to a temperature as a function of time.  Errors are propagated in the standard way.
%
%
%
%

\section{\label{sec:DLP}Double Langmuir probe}

Local plasma density and electron temperature measurements are performed using a double Langmuir probe (DLP) \cite{Johnson50}. A time series of ion saturation current from a DLP is a good proxy for electron density (see Figure \ref{BNdata}).  In the past, we have used a double probe on the SSX MHD wind tunnel to measure radial profiles of electron density and temperature, as well as local density fluctuations \cite{BrownPSST}.
%
%
%
%

A useful aspect of  a double Langmuir probe with identical electrodes is that the I-V characteristic is symmetric.  Since the entire circuit floats with the plasma, and if no potential difference is applied between the electrodes, the circuit will not extract any net current, i.e., $I(V=0) = 0$ (provided the plasma is quiescent).  Furthermore, since the current has the same magnitude (opposite sign) at $\pm V$, the maximum current is limited by the ion flux. Therefore, the I-V characteristic has the form  \cite{Johnson50}: 

$$I(V) =  I_{sat} \tanh \left( \frac{eV}{2kT_e} \right), $$ where the maximum current is given by:

$$I_{sat} = n e A \langle v \rangle = n e A \sqrt{\frac{k T_e}{M_i}}. $$  

Hence, the full expression can be written as:
\begin{equation}
I(V)=n e A \sqrt{\frac{k T_e}{M_i}}  \tanh \left( \frac{eV}{2kT_e} \right).
\label{tanh-fitting} 
\end{equation}
  
The SSX double Langmuir probe consists of two $1.5~mm$ diameter tungsten rods. The tungsten rods are installed in an alumina tube closed at the plasma facing end and cut from the sides so that only a nearly planar area of the probes is exposed to the plasma. The exposed probe areas are $1.5~mm$ long and $0.8~mm$ wide. The probe separation is about $3~ mm$. Probe tips were oriented across the flow direction to prevent one probe tip from shadowing the other.  

The probe tips are biased using a $360~\mu F$ capacitor bank charged with an external power supply that is isolated during the plasma discharge to prevent any ground loops. The voltage droop is typically less than $10\%$ after a discharge so the voltage between the probe tips is nearly constant.  The dynamical voltage difference between the probe tips is monitored using a Tektronix isolated voltage probe during a shot. A high bandwidth (100 MHz) current transformer (Tektronix TCP312A probe) reports the ion current flowing between the probe tips. Typical ion current magnitudes were $\leq 10~$A consistent with $I = n e v_{th} A$, and calibrated with a HeNe interferometer \cite{Buchenauer77,BrownPSST}. 

\begin{figure}[!h]
\begin{center}
\includegraphics[width=0.75\textwidth]{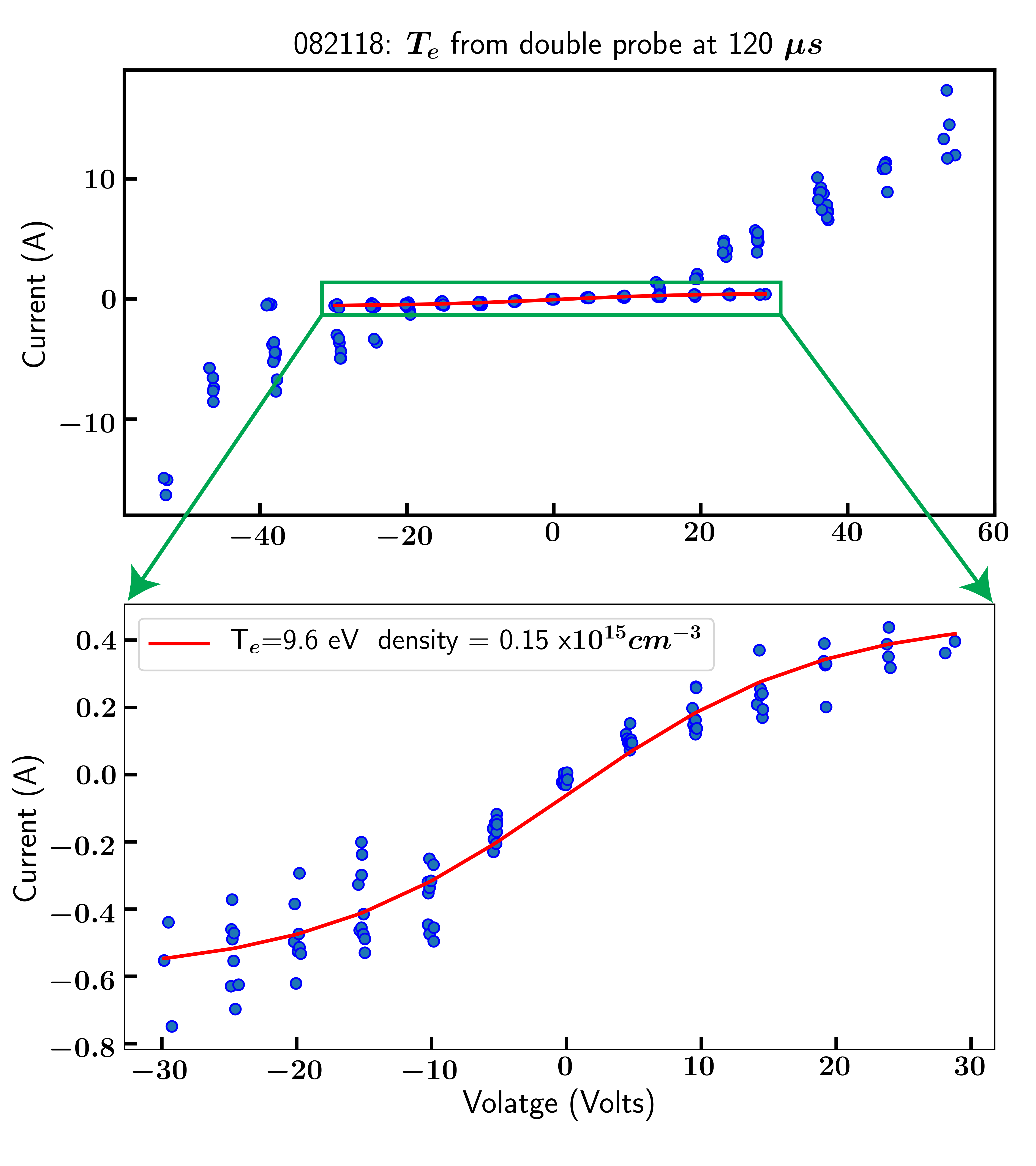}
\end{center}
\caption{Typical IV-curve of a double probe obtained using $\sim 130$ SSX shots.  This data was taken at $120~\mu s$ and averaged over a $10~\mu s$ window.  Data was fit to the tanh function Eq. \ref{tanh-fitting}.  The upper panel shows that at high bias voltages, secondary emission masks the probe characteristic.  The lower panel is the data selected for the fit.}
\label{IV-curve} 
\end{figure}

The biasing voltage between the probe tips was scanned from $-30~V$ to $+30~V$ with a $5~V$ interval and $10$ shots of the wind tunnel were recorded at each biasing voltage. One such scan is shown in Fig. \ref{IV-curve} at $120~\mu s$. Electron temperature and density are extracted from the double probe data at each time step using Eq. \ref{tanh-fitting}.  The fit yields $T_e = 9.6~eV$ and $n_e = 0.15 \times 10^{15}~cm^{-3}$ at $120~\mu~s$.  We suspect that the DLP is over-estimating $T_e$ (see Figure \ref{TeVUV}) but under-estimating density (see Figure \ref{density}) at $120~\mu s$, nonetheless we will continue to use it as a proxy for local density.

\section{\label{sec:Results}Experiment and Results}

\begin{figure}[!h]
\begin{center}
\includegraphics[height=3.5in]{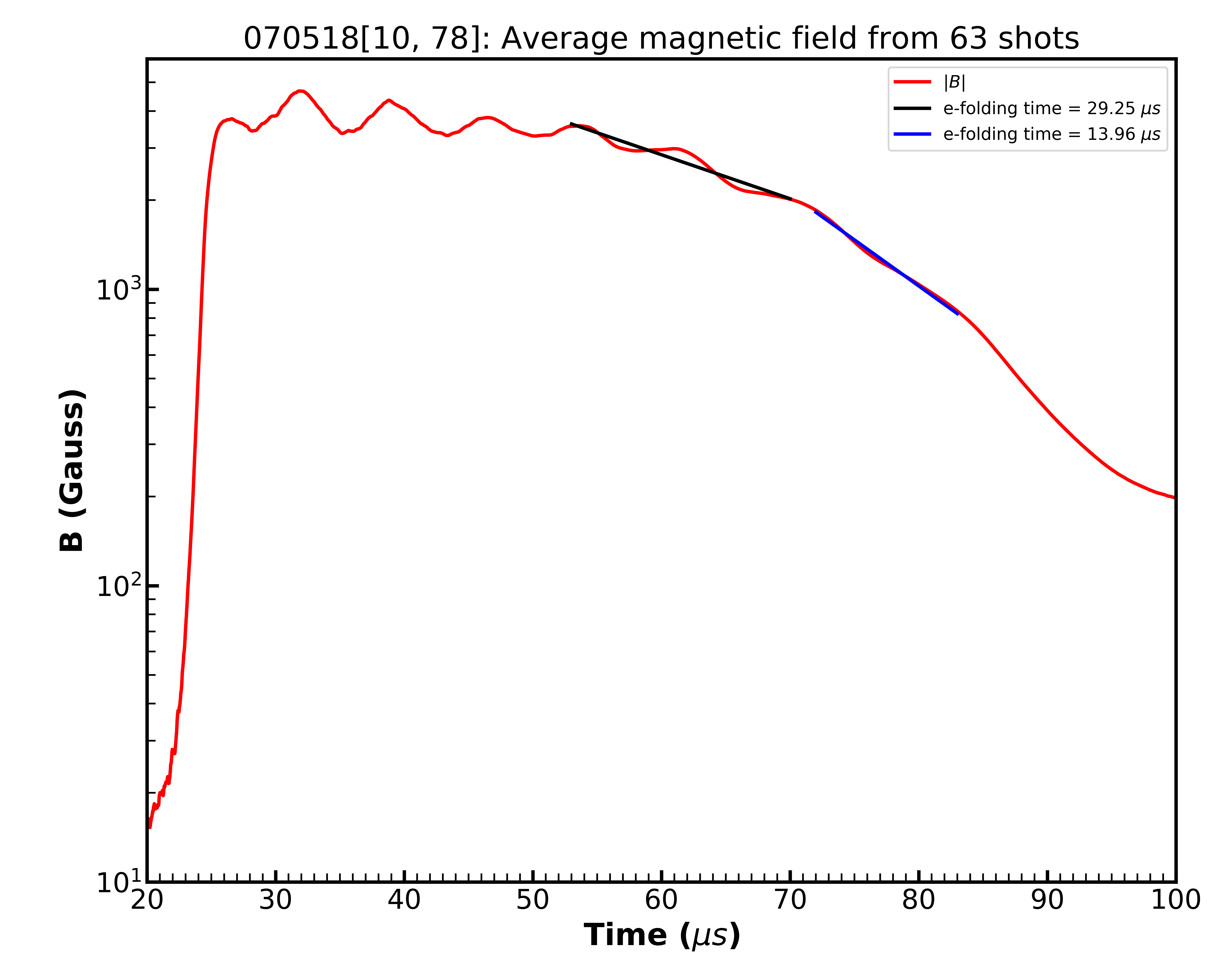}
\end{center}
\caption{Time trace of average magnetic field $|B|$ obtained from the small $\dot{B}$ probe, at $25~cm$ away from the gun, measuring three orthogonal components of magnetic field. The average is obtained using 63 shots taken under similar conditions. The average magnetic field shows an e-folding time of $30~\mu s$ (shown by the black straight line) at earlier time whereas at later time, the e-folding time is around $14~\mu s$ (shown by the blue straight line).}
\label{B_Lifetime} 
\end{figure}

We found in Section \ref{sec:VUV2} that our electron temperature is about $7~eV$ for most of the discharge.  Using our model (Equation (1)), we can calculate an e-folding lifetime of $\tau = \frac{\mu_0}{\lambda^2 \eta} = 29~\mu s.$ In Figure \ref{B_Lifetime}, we show the average magnetic field $|B|$ as a function of time. If we extract an e-folding time as the Taylor state begins to decay, we find $\tau = 30~\mu s$.  Later in time, the decay is faster.
%
%

\begin{figure}[!h]
\begin{center}
\includegraphics[height=3.5in]{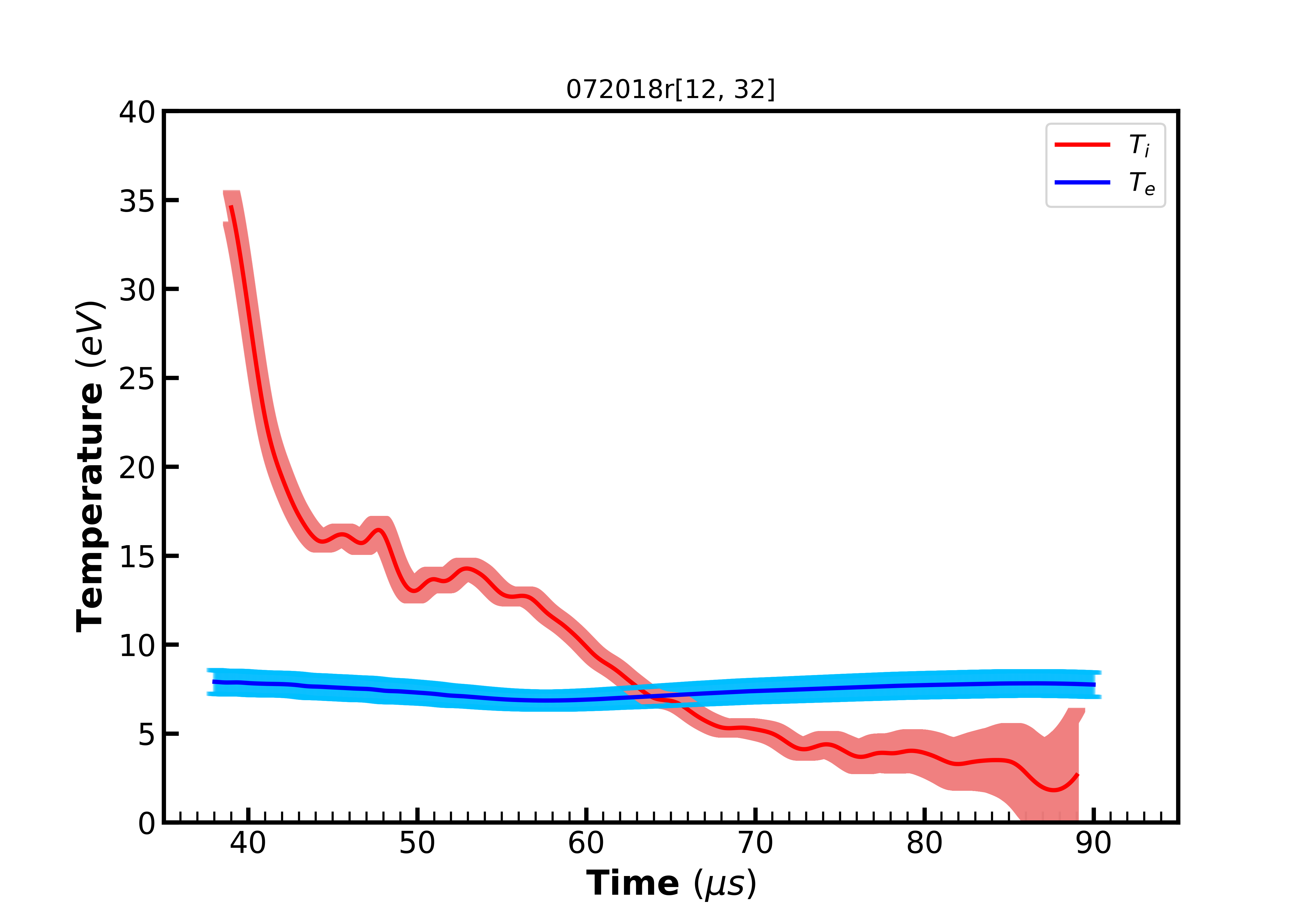}
\end{center}
\caption{Time trace of average ion temperature (measured using ion Doppler spectroscopy, $124~cm$ away from the gun) and electron temperature obtained from VUV spectroscopy measured $5~cm$ away from the gun.}
\label{Temp} 
\end{figure}

In addition, we can measure the proton temperature using ion Doppler spectroscopy \cite{CothranRSI}.  We find that the proton temperature in SSX is initially always higher than $T_e$, consistent with expectations for a magnetized coaxial gun (Marshall gun).  In Figure \ref{Temp} we show the cooling of warm protons (measured downstream of the gun) and superpose our $T_e$ result.  Although they are measured at different locations (124 cm away from the gun for the protons, 5 cm for the electrons), we can see if the warm protons could be cooling on the electrons.  Since electron heat flux is approximately: $Q_e \approx 0.71 nkT_e v_{\parallel}$ and $v_{\parallel} \approx 1~m/\mu~s$, the electrons are in local thermal equilibrium during the $100~\mu s$ evolution of the experiment. We don't expect gradients in $T_e$ along field lines.  From our model (Equation (2)),  we calculate an e-folding cooling time of $\sim 1~\mu s$ (using $n_e = 5 \times 10^{15}$ and $7~eV$). Evidently, other effects are preventing the protons from cooling rapidly.  We suspect that energy from the MHD cascade ultimately puts heat into the ion component \cite{Maruca11}.
%
%
%
%
%
%
%
%
 
\section{\label{sec:Summary}Discussion and Summary}

High ion temperature is an obvious requirement for practical fusion energy since fusion cross-sections peak at temperatures over $10~keV$.  In addition, the Lawson criterion requires density, temperature, and confinement time such that $n T \tau \ge 10^{21}~keV s~m^{-3}$.  However, it is also critical that electron temperatures are also high since ions will eventually cool on electrons as our results in Section \ref{sec:Results} suggest. It is particularly important for MIF schemes since magnetic lifetime scales like $\tau \propto T_e^{3/2}$ (Equation 1).

\begin{figure}[!h]
\begin{center}
\includegraphics[height=3.5in]{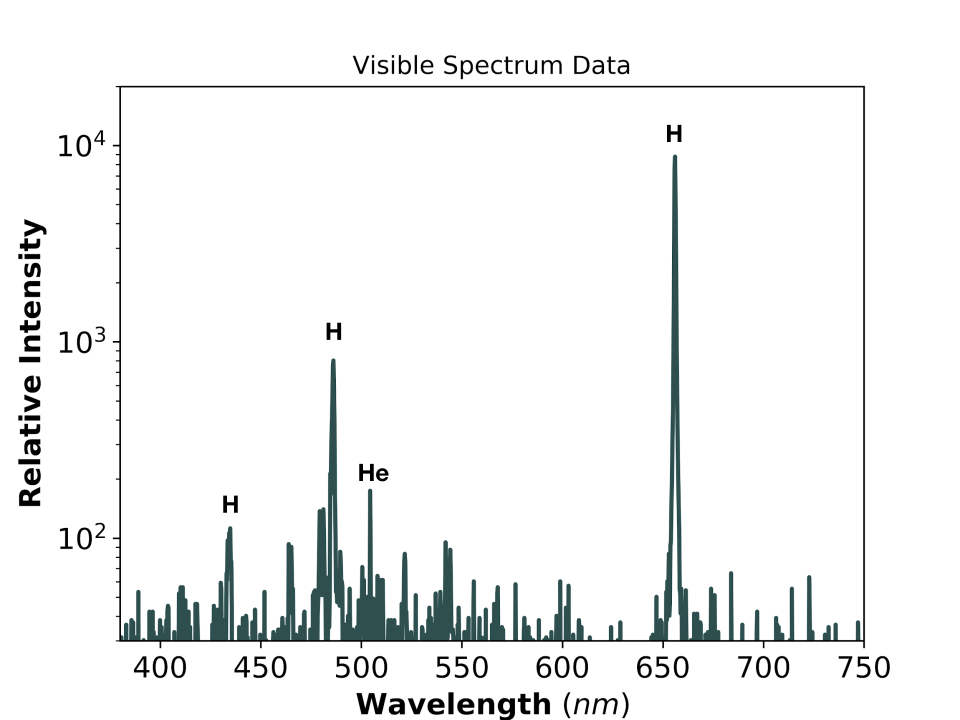}
\end{center}
\caption{Visible Spectrum produced by our Ocean Optics Spectrometer, with the strongest lines labeled with the most likely atomic source. Multiple Hydrogen lines are present  - including  H$\alpha$ at $656~nm$,  H$\beta$ at $486~nm$, and H$\gamma$ at $434~nm$. There is also a Helium line present around $501~nm$, which indicates that our plasma was contaminated by some amount of Helium. }
\label{Visible} 
\end{figure}

In order for a Taylor state to be a suitable MIF target, we need to work to increase $T_e$.  One approach is to maintain high vacuum conditions and clean plasma facing walls.  We have a meticulous process for cleaning the vacuum walls of SSX involving baking the stainless steel vacuum chamber to $250^o~C$ and running a Helium glow discharge in the wind tunnel section.  In any case, we find that there are still some impurities.  In Figure \ref{Visible} we show a visible spectrum from an Ocean Optics visible spectrometer, averaged over an entire shot.  It shows very bright recombination lines from hydrogen (as expected) but also some residual Helium lines from the glow discharge cleaning, as well as several other unidentified lines.  Since the wind tunnel is coated with plasma-sprayed tungsten, we suspect that some gas is trapped in the interstices of the tungsten. To improve our Taylor state lifetimes, more work is needed to rank and mitigate the relevant cooling mechanisms for electrons.

\section{\label{sec:Ack}Acknowledgements} This work was supported by the DOE Advanced Projects Research Agency (ARPA) ALPHA program project DE-AR0000564.  The authors wish to acknowledge the support and encouragement of ARPA program manager Dr. Patrick McGrath.  Technical support from Steve Palmer and Paul Jacobs at Swarthmore for SSX is also gratefully acknowledged.

\bigskip

\newpage

\end{document}